\documentclass[conference]{IEEEtran}

\IEEEoverridecommandlockouts

\usepackage{cite}
\usepackage{amsmath,amssymb,amsfonts}
\usepackage{algorithmic}
\usepackage{booktabs}
\usepackage[utf8]{inputenc}
\usepackage{graphicx}
\usepackage{pgfplots}
\usepackage{textcomp}
\usepackage{tikz}
\usepackage{xcolor}
\usepackage{hyperref}
\usepackage[normalem]{ulem}
\usepackage{amsthm}

\DeclareUnicodeCharacter{2212}{−}
\usepgfplotslibrary[groupplots,dateplot]
\usetikzlibrary[patterns,shapes.arrows]
\pgfplotsset{compat=newest}

\usepackage[capitalise]{cleveref}

\theoremstyle{definition}

\newcommand{\bB}{\mathbf{B}}
\newcommand{\be}{\mathbf{e}}

\newcommand{\bL}{\mathbf{L}}

\newcommand{\bg}{\mathbf{g}}

\newcommand{\im}{\text{im}}
\newcommand{\bH}{\mathbf{H}}
\newcommand{\bh}{\mathbf{h}}
\newcommand{\bef}{\mathbf{f}}

\newcommand{\bc}{\mathbf{c}}

\begin{document}

\title{
    Outlier Detection for Trajectories via Flow-embeddings
    \thanks{*The first two authors contributed equally to this work. FF and MTS acknowledge funding from by the Ministry of Culture and Science (MKW) of the German State of North Rhine-Westphalia (``NRW Rückkehrprogramm'') and the Excellence Strategy of the Federal Government and the Länder }
}

\author{
    \IEEEauthorblockN{Florian Frantzen$^*$}
    \IEEEauthorblockA{
        \textit{Department of Computer Science} \\
        \textit{RWTH Aachen University}\\
        Aachen, Germany \\
        florian.frantzen@cs.rwth-aachen.de
    }
    \and
    \IEEEauthorblockN{Jean-Baptiste Seby$^*$}
    \IEEEauthorblockA{
        \textit{Operations Research Center} \\
        \textit{Massachusetts Institute of Technology}\\
        Cambridge, MA, USA \\
        jbseby@mit.edu
    }
    \and
    \IEEEauthorblockN{Michael T. Schaub}
    \IEEEauthorblockA{
        \textit{Department of Computer Science} \\
        \textit{RWTH Aachen University}\\
        Aachen, Germany \\
schaub@cs.rwth-aachen.de    }
}

\maketitle

\begin{abstract}
    We propose a method to detect outliers in empirically observed trajectories on a discrete or discretized manifold modeled by a simplicial complex.
    Our approach is similar to spectral embeddings such as diffusion-maps and Laplacian eigenmaps, that construct vertex embeddings from the eigenvectors of the graph Laplacian associated with low eigenvalues.
    Here we consider trajectories as edge-flow vectors defined on a simplicial complex, a higher-order generalization of graphs, and use the Hodge 1-Laplacian of the simplicial complex to derive embeddings of these edge-flows.
    By projecting trajectory vectors onto the eigenspace of the Hodge 1-Laplacian associated to small eigenvalues, we can characterize the behavior of the trajectories relative to the homology of the complex, which corresponds to holes in the underlying space.
    This enables us to classify trajectories based on simply interpretable, low-dimensional statistics.
    We show how this technique can single out trajectories that behave (topologically) different compared to typical trajectories, and illustrate the performance of our approach with both synthetic and empirical data.
\end{abstract}


\section{Introduction}
Trajectory data on discrete or continuous spaces appears in many applications such as autonomous systems, self-driving cars \cite{Xu2012}, GPS tracking of objects \cite{Sun2019}, animal and humans migrations \cite{Thums2018}, physical experiments that study particles or fluid trajectories \cite{VanDenBremer2019}, or digital traces on the Web \cite{Hepp2018}. 
The goal of this paper is to introduce a new method of outlier detection for such trajectories on discrete (or discretized) spaces based on discrete Hodge Theory.

Outlier detection is a prevalent task that appears in pre-processing data, or for related tasks such as clustering, hypothesis testing and change-point detection~\cite{zheng2015,Meng2018}.
The key modeling question underpinning this task is how to define and model valid data, as opposed to an ``outlier''.
Often, this involves defining a suitable representation of a trajectory (e.g., measuring certain features of a trajectory), based on which one can then measure a distance of the trajectory of interest to some assumed normal behavior.
In the context of trajectories on discrete spaces, graphs are typically used to model the underlying space and trajectories then simply amount to walks (flows) on such graphs.

In contrast, here we employ simplicial complexes (SCs), which can be viewed as an extension of graphs rooted in algebraic topology, to model the underlying space.
This is motivated by the notion of outlier we are interested in: specifically, we say a trajectory is an outlier if it takes a ``topological detour'', i.e., behaves differently with respect to certain obstacles or landmarks within the underlying (discrete) space.
We model these obstacles as ``holes'' in the underlying space, which can be appropriately encoded in an SC.
Using spectral features of the $1$-Hodge-Laplacian associated to this SC we then derive a simple spectral embedding of the trajectories, which we can employ to detect outliers using standard methods for vector space data.

\noindent\textbf{Related literature}
SCs and their spectral properties in terms of the Hodge-Laplacian have been studied in \cite{Lim2015,schaub2020random,Barbarossa2018}.
Specifially,~\cite{Barbarossa2020,Schaub2018a,yang2021finite} develop methods for flow estimation and filtering,~\cite{Jia2019} addresses semi-supervised and active learning, and~\cite{schaub2021} review signal processing techniques for SCs and hypergraphs. 
Following initial work on edge-flow embeddings in~\cite{schaub2020random}, our goal here is to develop a systematic method to detect outliers among trajectories.
Of particular relevance is the work on trajectory prediction that has been recently studied in \cite{Roddenberry2021,ghosh2018topological}.

\noindent\textbf{Outline}
The remainder of this paper is structured as follows:
In \cref{section:background}, we briefly explain essentials of algebraic topology.
In \cref{section:method}, we describe our outlier detection method and provide a proof of concept, before provide a more thorough numerical study using synthetic data in \cref{section:synthetic_experiment}.
In \cref{section:realData}, we apply our method to a real data set of whale trajectories.
Finally, \cref{section:discussion} provides a discussion of our method, concluding remarks and directions for future work.

\section{Background in algebraic topology}\label{section:background}
In this section, we present an elementary overview of the concepts from algebraic topology used to process signals defined on SCs.
For more details, see~\cite{schaub2021}.

\noindent\textbf{Simplicial complexes and edge-flows}
Given a set of vertices $\mathcal{V}$, a $k$-simplex $\mathcal{S}^k$ is a subset of $k+1$ vertices.
An abstract simplicial complex (SC) $\mathcal{X}$ is a collection of simplices, such that for any $k$-simplex $\mathcal{S}^k$ in $\mathcal{X}$, any subset of $\mathcal{S}^k$ is also in $\mathcal{X}$.
In the following we will focus on SCs containing $0$-simplices, $1$-simplices, and $2$-simplices, which we call \emph{vertices}, \emph{edges}, and \emph{triadic faces}, in analogy to graphs~\cite{schaub2021}.
The sets of vertices and edges are denoted with $\mathcal{V}$ and $\mathcal{E}$, respectively.

\begin{figure}
    \centering

    \begin{minipage}{0.40 \linewidth}
        \includegraphics[width=\linewidth]{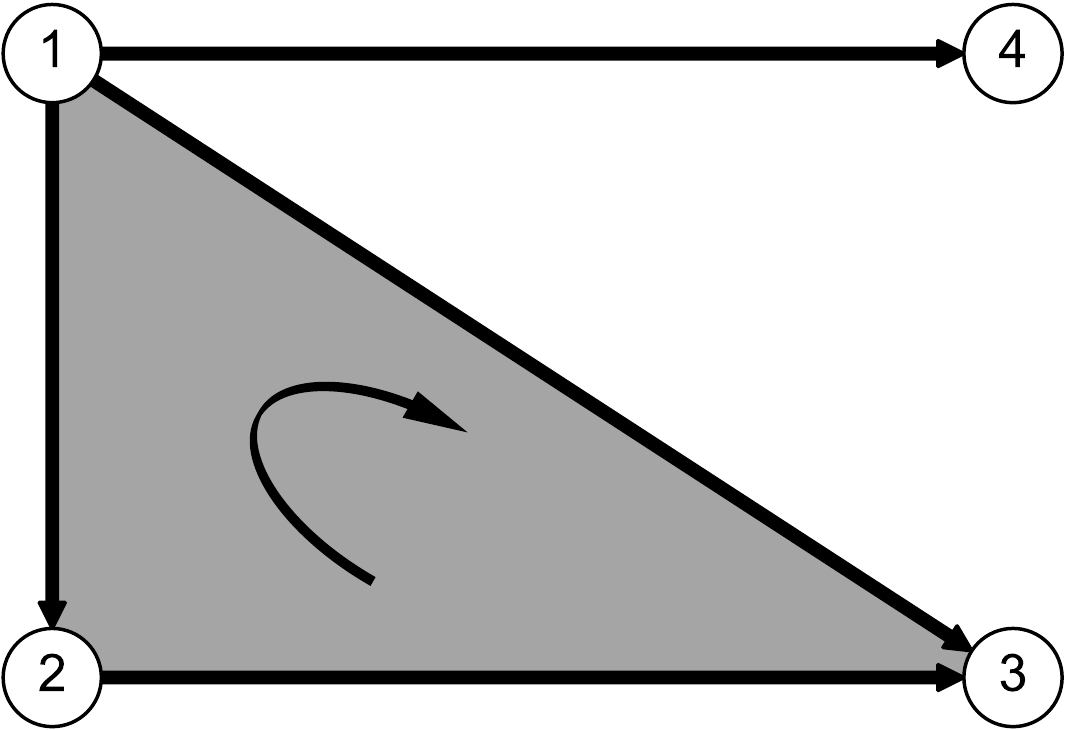}
    \end{minipage}
    \hfill
    \begin{minipage}{0.55 \linewidth}
        \scriptsize
        \let\quad\thinspace 
        \begin{align*}
              \bordermatrix{
                & (1,2)&(1,3)&(1,4)&(2,3)\cr
                1&1&1&1&0\cr
                2&-1&0&0&1\cr
                3&0&-1&0&-1\cr
                4&0&0&-1&0
            }=\mathbf{B}_1 \\
                  \bordermatrix{
                & (1,3,2)\cr
                    (1,2)&-1\cr
                    (1,3)&1\cr
                    (1,4)&0\cr
                    (2,3)&-1\cr
                }=\mathbf{B}_2
                \end{align*}
    \end{minipage}

    \caption{
        \textbf{Illustration of a simplicial complex and its boundary operators.}
        The SC has a filled triangle in shaded area.
        The direction of the arrows indicate the (arbitrary) chosen reference orientation.
    }
    \label{fig:toy_example}
\end{figure}

We encode the structure of an SC via incidence matrices $\bB_k$, which record the relations between $(k-1)$-simplices and $k$-simplices.
Note that $\bB_k$ are simply the matrix representations of the boundary operators.
More explicitly, rows of $\mathbf{B}_k$ are indexed by $(k-1)$-simplices and columns of $\mathbf{B}_k$ are indexed by $k$-simplices. 
Thus, the matrix $\mathbf{B}_1$ is the vertex-to-edge incidence matrix and $\mathbf{B}_2$ is the edge-to-triangle incidence matrix.
As a matter of book-keeping, we number the vertices in the SC from $1$ to $N$ and define a reference orientation for each simplex by increasing ordering of its vertices.
Thus, each edge $\mathbf{e}$ has an arbitrary orientation from its tail $t(\be)$ to its head $h(\be)$.
The entries of $\bB_1$ are then defined as $(\bB_1)_{t(\be),\be} = - 1, (\bB_1)_{h(\be),\be} = 1$ and $(\bB_1)_{k,\be} = 0$ otherwise.
Similarly, for each (oriented) triadic face $\Delta$ delimited by a triplet of edges $\{\be_1, \be_2,\be_3\}$, we have $(\bB_2)_{\be_i, \Delta} = 1$ if $\be_i$ is aligned with the reference orientation of the triangle, $(\bB_2)_{\be_i, \Delta} = -1$ if $\be_i$ it is not aligned. Further, $(\bB_2)_{\be_i, \Delta} = 0$ if $\be_i$ is not incident to the triadic face.
An illustration of these constructions is given in~\cref{fig:toy_example}.

We define an edge-flow (edge-signal) on an SC as a real-valued function $f \colon \mathcal{V} \times \mathcal{V} \rightarrow \mathbb{R}$ such that:
\begin{align}
    f(i,j) = \begin{cases}
        -f(j,i), & \text{if } (i, j) \in \mathcal{E} \\
        0,       & \text{otherwise}.
    \end{cases}
\end{align}
Since, we consider SCs with a finite number of edges, we can represent any such edge flow by a vector $\mathbf{f}$, indexed by the edges in the chosen reference orientation.

\noindent\textbf{The Hodge Laplacian and the Hodge Decomposition}
Using the above incidence matrices, we can define the Hodge {$1$-Laplacian} as $\textbf{L}_1 = \mathbf{B}_{1}^\top \mathbf{B}_{1} + \mathbf{B}_{2} \mathbf{B}_{2}^\top$. 
The eigenvectors of this operator give rise to the \emph{Hodge decomposition}, which states that the space of edge flows, which is isomorphic to $\mathbb{R}^E$, can be decomposed as:
\begin{align}\label{eq:Hodge_decomposition}
    \mathbb{R}^{E} =  \im(\bB_{2})\oplus \im(\bB_1^\top) \oplus \ker(\bL_1),
\end{align}
where $\im(\bB_{2})$ is the space of curl flows, $\im(\bB_1^\top)$ is the space of gradient flows, and $\ker(\bL_1)$ is the space of harmonic flows.
This decomposition enables us to decompose any edge flow $\bf$ into a curl $\bc \in \im(\bB_{2})$, a gradient $\bg \in \im(\bB_1^\top)$, and a harmonic flow $\bh \in \ker(\bL_1)$.
Importantly, the space of harmonic flows is isomorphic to the homology group of the underlying SC, i.e., encodes the ``holes'' in underlying space~\cite{Lim2015,schaub2020random,schaub2021}. 

\section{Method}\label{section:method}
In the following, we describe our outlier detection method, by means of a concrete example. 
At an intuitive level, the methods may be described as follows:
We model the (discrete) space on which the trajectories take place by means of an SC with a nontrivial homology, i.e., with ``holes''. 
We are interested in how the trajectories behave relative to these holes.
We therefore construct the Hodge-Laplacian of the SC and compute the associated harmonic flows, which correspond to global circular flows around the holes
~\cite{schaub2020random,schaub2021}.
We then represent each trajectory as the indicator vector (of an edge-flow) which we project onto the harmonic flows, thus creating a low-dimensional embedding into a Euclidean vector space that describes the behavior of the trajectory.
Within this vector space we can then detect outliers using standard methods for outlier detection for point cloud data.

Obviously, this approach requires the existence of holes in the SC, i.e., particular obstacles or landmarks in the space we are interested in.
In the following we assume that these landmarks are given already.
In general, such landmarks may be inherently defined by the problem itself, e.g., ships cannot move through islands, or may be decided upon by the researcher, i.e., the researcher may actively place certain landmarks (in terms of holes) into the complex, he is particularly interested in.
Landmarks can also be informed by the data if certain areas are not traversed by any trajectories.

\begin{figure}
    \centering

    \includegraphics[width=\linewidth]{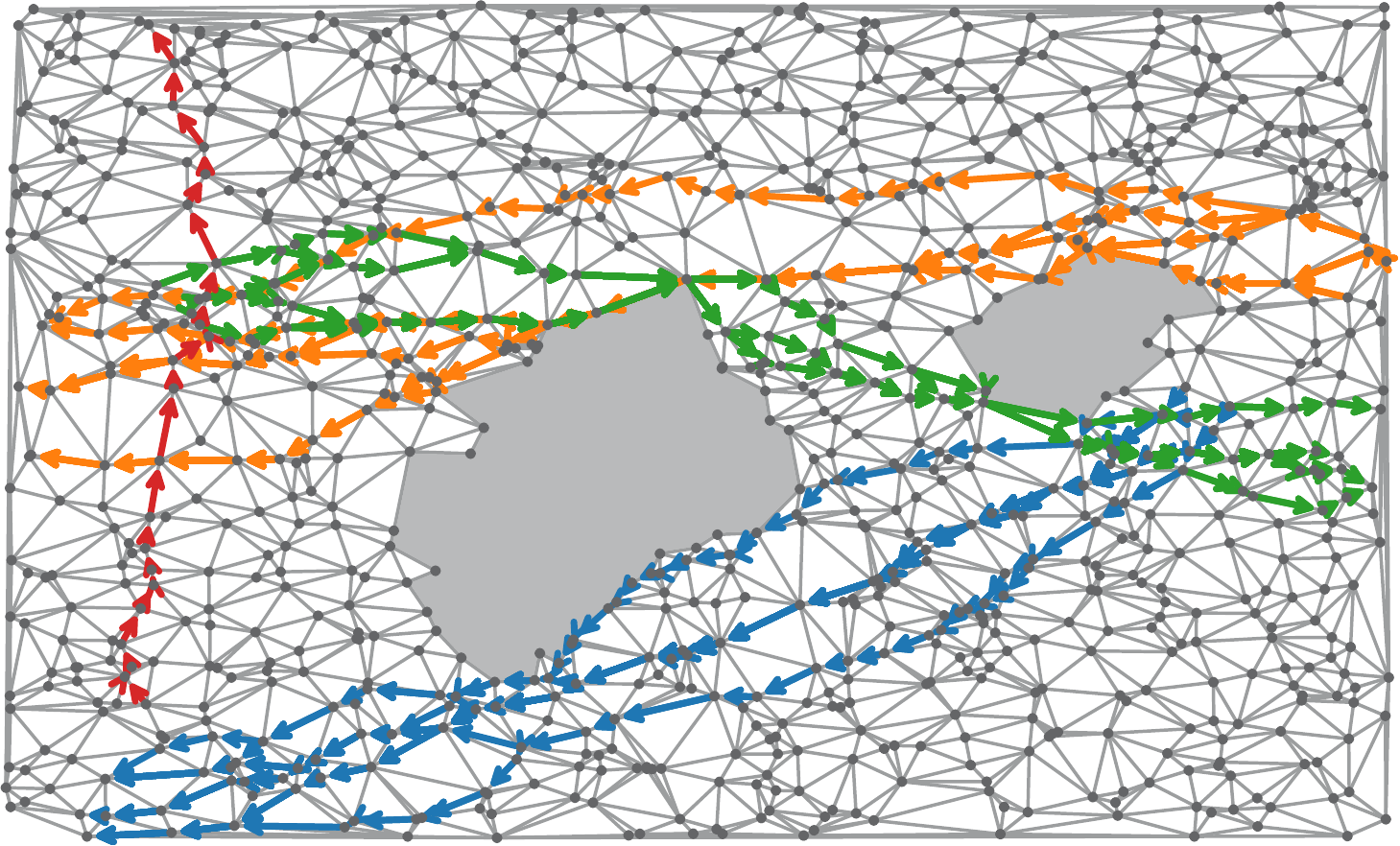}

    \caption{
        \textbf{Illustration: Trajectories on a simplicial complex.}
        We construct an SC with two holes (shaded areas) using Delaunay triangulation.
        On this SC, we define three classes of trajectories (blue, green and orange) of 40 similar trajectories each.
        An additional outlier trajectory (red) is planted on the SC.
    }
    \label{fig:Delaunay_map}
\end{figure}

Let us now illustrate these idea by means of a concrete example.
We construct an SC by drawing $1000$ points uniformly at random in the unit square and construct a triangular lattice using Delaunay triangulation.
We consider all triangles as $2$-simplices, i.e., triadic faces.
We then remove some vertices and all their adjacent edges to add two ``holes'' (landmarks) to the SC.
The resulting SC is displayed in \cref{fig:Delaunay_map}, with the two holes shown as shaded areas.
Accordingly, the Hodge Laplacian $\bL_1$ has two zero eigenvalues (corresponding to the two holes).

We construct three synthetic classes of trajectories that behave differently with respect to the holes in the SC (see~\Cref{fig:Delaunay_map}).
Each class corresponds to one departure and one arrival area and contains $40$ trajectories. 
For each class the first trajectory is created as the shortest path between the departure and the arrival vertex.
To create subsequent trajectories, we increase the edge weights along the previously found path and compute the shortest path on the adjusted SC, leading to a new trajectory.
In addition to these normal trajectory classes, we add one outlier trajectory (see~\cref{fig:Delaunay_map}).

\begin{figure}[t]
    \centering
    \includegraphics[width=\linewidth]{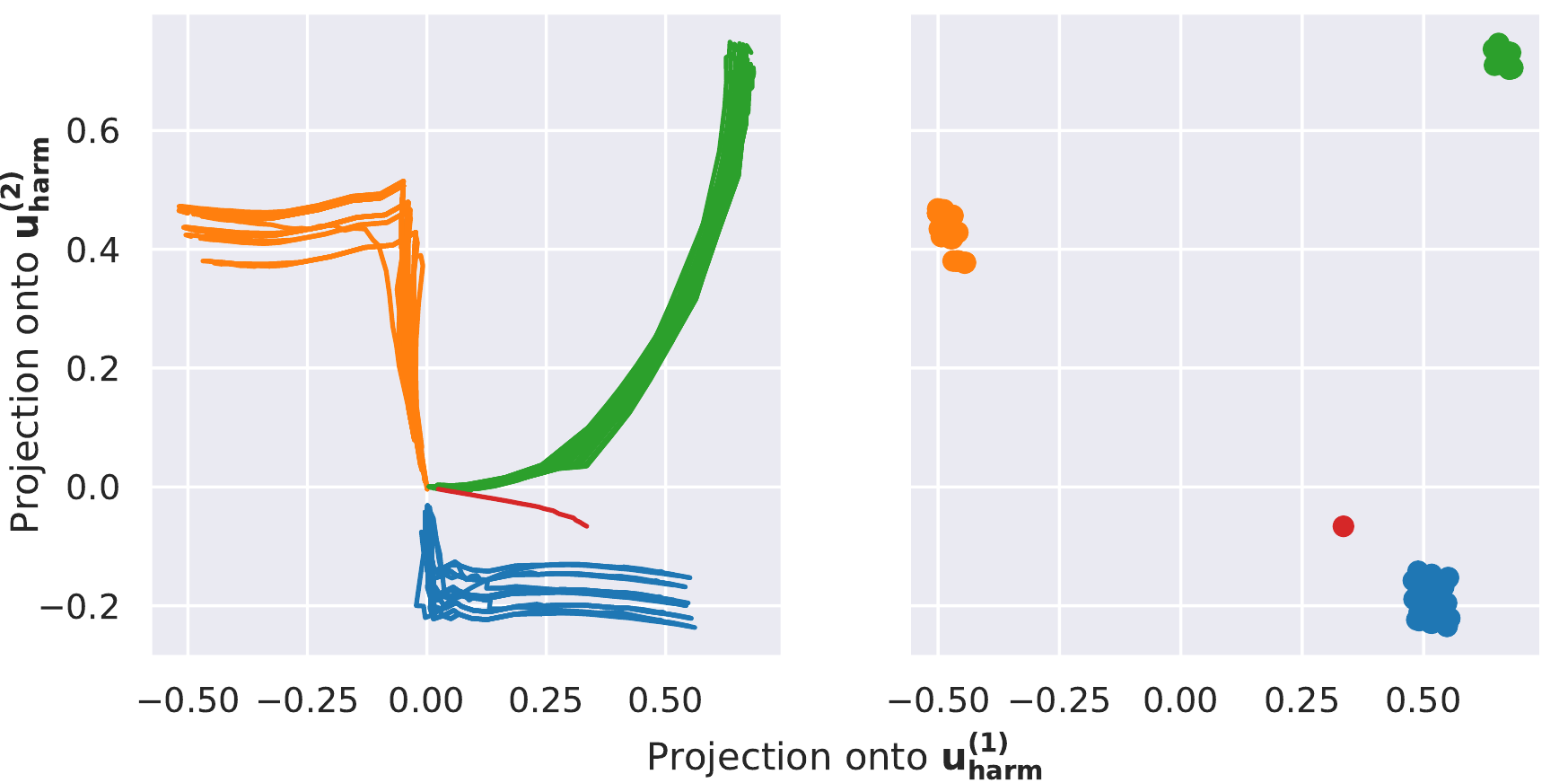}
    \caption{
        \textbf{Harmonic trajectory embeddings for our example simplicial complex.}
        The left plot shows the embedding for one edge at a time.
        On the right, the flattened trajectory embeddings are plotted.
    }
    \label{fig:embedding-space}
\end{figure}

We then project each trajectory onto the harmonic space.
A trajectory can be represented by a vector $\bef$ with entries $f_{[i,j]} = 1$ if the edge $[i,j]$ is traversed in the chosen reference orientation, $f_{[i,j]} = -1$ if the edge is traversed in the opposite orientation, and $0$ otherwise.
Let us assemble the harmonic eigenvectors of the Hodge-Laplacian in the $\mathbf{H}$.
Using the mapping $\mathbf{e} \mapsto  \mathbf{H}^\top \mathbf{e}$, each single edge $\mathbf{e}$ traversed by the trajectory can then be represented by a point with coordinates $(u_1, u_2)$ in the harmonic space spanned by the eigenvectors $\bf{u_{harm}^{(1)}}$ and $\bf{u_{harm}^{(2)}}$ with eigenvalue zero.
The result is displayed in \cref{fig:embedding-space}\textbf{A}. 
Successively projecting each edge that composes a trajectory onto the harmonic space and joining the projection of consecutive points by a segment, we obtain a trajectory in the embedding space. 
This embedding is shown in Figure \ref{fig:embedding-space}\textbf{A}.
Alternatively, we can simpy embed the full trajectory by a single point (\cref{fig:embedding-space}\textbf{B}).
An embedding of the full trajectory is obtained via the mapping $\bef \mapsto \mathbf{H}^\top \mathbf{f}$. 
In the remainder of this paper, $\bef$ is called the \emph{flattened trajectory} vector and  $\mathbf{H}^\top \mathbf{f}$ is the corresponding \emph{flattened embedding}.

\begin{figure}[t]
    \centering
    \includegraphics[width=\linewidth]{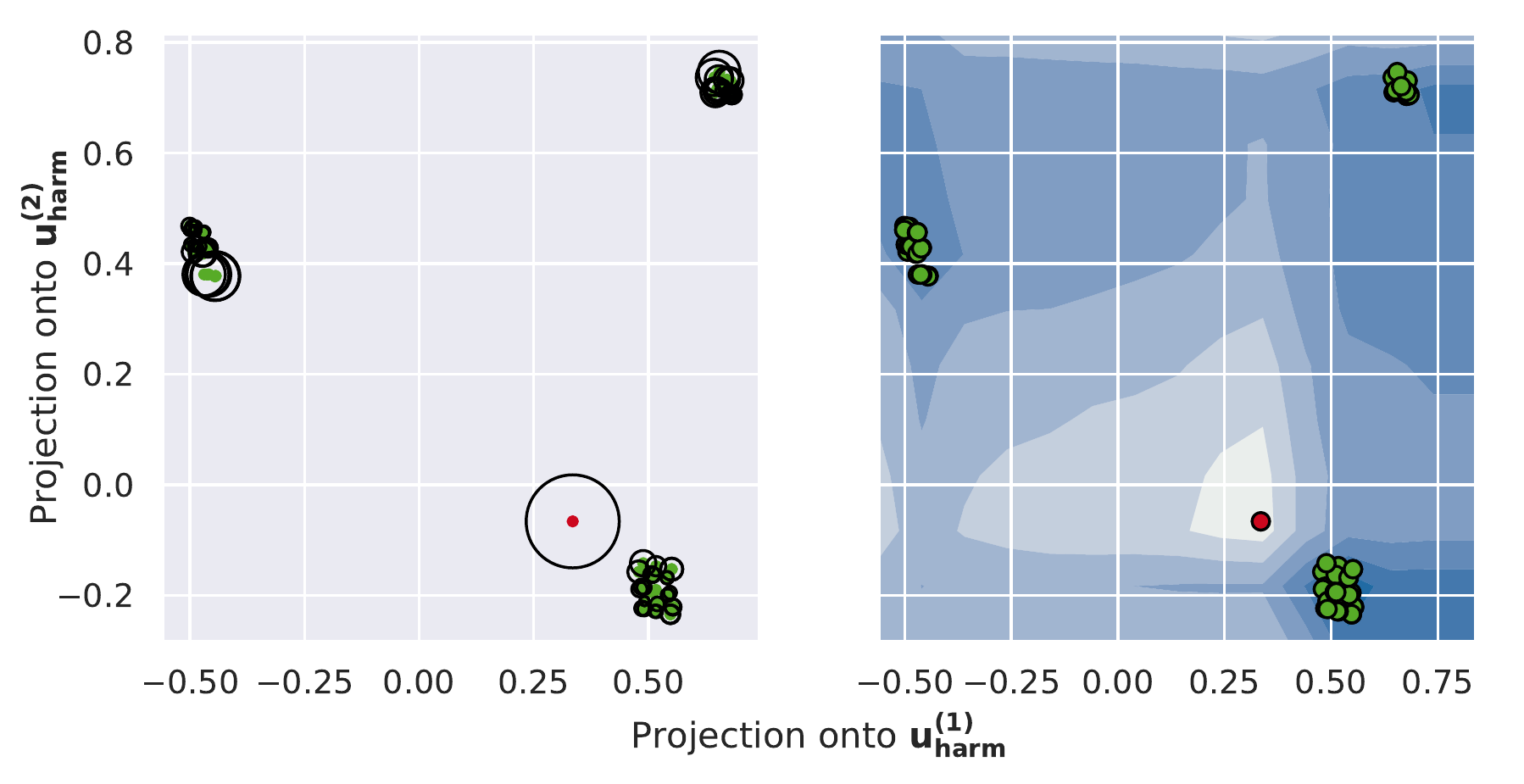}
    \caption{
        \textbf{Outlier detection via local outlier factor (left) and isolation forest (right) applied to the harmonic embeddings.}
        For LOF, the outlier factors for each point is plotted as a circle, where a greater radius corresponds to a larger outlier factor.
        For IS, the background contour shows the decision function, where lighter colors correspond to larger anomaly scores.
        In both cases, the planted outlier trajectory is predicted correctly.
    }
    \label{fig:outlier-predictions}
\end{figure}

We observe that trajectories within the same class have a similar representation in the embedding space, which enables us to detect the outlier data via standard tools for outlier detection for Euclidean data.
We use two existing algorithms: Local Outlier Factor (LOF) \cite{breunig2000lof} and Isolation Forest (IF) \cite{liu2008isolation}.
Results of the outlier detection using these methods are shown in \cref{fig:outlier-predictions}. 
For both methods, three main clusters corresponding to the three classes are identified and the outlier trajectory is isolated into its own cluster. 

\section{Numerical experiments with synthetic data}
\label{section:synthetic_experiment}

In this section, we assess the performance of our outlier detection approach using synthetic data created as described in the last section.
Specifically, we generate $100$ distinct synthetic data sets according to the previously outlined method.
Within each dataset, we plant five random outlier trajectories.

Notice that this method is rather naive in terms of generating similar trajectories for each class: It frequently happens that the planted obstacle is directly in the way and the shortest paths sometimes go left and sometimes go right around the obstacle.
This effectively splits the trajectory class into two or more classes in our interpretation, leading to more than the desired three classes.
However, as long as there are sufficiently many trajectories for each split, we ignore this problem in our data generation process (if there is a single trajectory in such a split, it would be considered an outlier).
Indeed, we are only interested in classifying the outliers and not in recovering the trajectory classes. 
Hence, we do not need the ground-truth class labels to discriminate the outliers from the valid trajectory classes.

We compute the Hodge Laplacian for each artificially created SC and embed the flattened trajectory vectors into the corresponding two-dimensional harmonic space.
On the resulting embeddings, we use local outlier factors and isolation forests with contamination parameter $0.04$ to detect anomalies in the data.
Computing the prediction metrics for this setup, we observe an accuracy of $0.9754$ when using local outlier factors and an accuracy of $0.9892$ using isolation forests.
We remark that the performance of both methods could potentially be further improved, e.g., by optimizing the embedding procedure; we postpone those aspects to future work.

\section{Numerical experiments with real data}
\label{section:realData}

We further illustrate the performance of our approach on some real-world trajectory data.
For that, we use a dataset of 41 whale trajectories around the Canadian Arctic Archipelago\footnote{The data is available under \url{https://www.movebank.org/cms/webapp?gwt_fragment=page=studies,path=study467034665}. All code to reproduce the here presented experiments is available at \url{https://git.rwth-aachen.de/netsci/trajectory-outlier-detection-flow-embeddings}}.
The Canadian Arctic Archipelago of big and small islands, around which the whales may swim, enables a natural construction of obstacles for the underlying space we model again by an SC.

To construct an SC, we follow the approach of~\cite{schaub2020random} and discretize the earth's surface using a hexagonal grid, with the width of each hexagon corresponding to 0.86° (latitude).
We associate a vertex with each hexagon and connect two vertices by an edge if their hexagons have a common face.
Every triplet of hexagons that meet at a common point form a filled-in triangle.
Hexagons that only cover landmasses are removed, which leads to some holes (obstacles) in the SC: clearly the whale trajectories cannot pass through land.
Finally, we discretize the whale trajectories by rounding their positional coordinates to the nearest hexagon centers, while dropping any consecutive locations in the same hexagon.
Due to missing data, trajectories can sometimes ``jump'' from one hexagon to some non-neighboring hexagon.
In these cases, we fill in the missing data by calculating a shortest path between both endpoints on the SC, ensuring that all trajectories are continuous.
We consider the resulting sequence of edges that the trajectory traverses in the SC as the discretized trajectory of a whale.

\begin{figure}
    \centering
    \includegraphics[width=\linewidth]{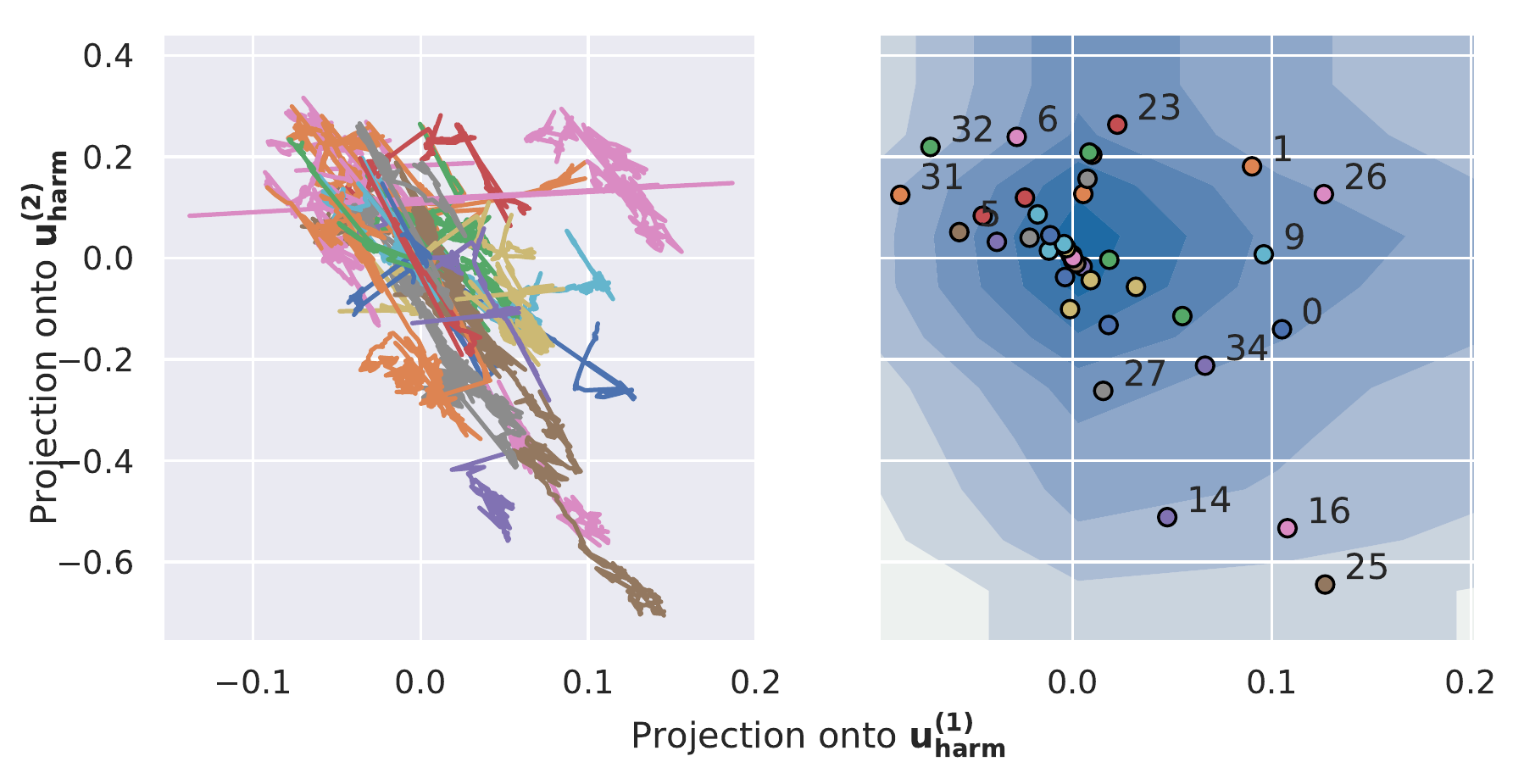}
    \caption{
        \textbf{Harmonic embeddings and outliers of whale trajectories.}
        The left figure shows the full trajectory embeddings and the right figure the corresponding flatten embeddings.
        The contour illustrates the decision function of an isolation forest fitted on the point cloud.
        Observe a high concentration in a defined area, some clear outliers $14$, $16$ and $25$ in the bottom right and some possible outliers at the boundaries of the center.
    }
    \label{fig:whale-predictions}
\end{figure}

Given this SC and the discretized trajectories supported on it, we can compute harmonic embeddings for the whales as before:
We construct the Hodge Laplacian $\bL_1$ of the SC, compute the harmonic eigenvectors (corresponding to zero eigenvalues, in this case there are again 2) and use them to construct a harmonic projection matrix $\bH$.
For each trajectory, we compute the full and the flattened harmonic embeddings, as can be seen in \cref{fig:whale-predictions}.
Even though the full embedding contains significantly more information, for simplicity we limit ourselves to only use the simpler flattened embeddings.
Empirically, we find a high concentration of most embedding points in an area around the origin of the embedding space.
We again use an isolation forest to find the outliers in the resulting point cloud.
The estimator predicts the trajectories $14$, $16$ and $25$ as clear outliers
(depending on the chosen threshold there are some further \textit{possible} outliers).

Since the dataset does not contain a ground-truth for outliers, we can only give a qualitative reasoning as to why the predicted outliers are a sensible choice.
For that we compare the identified outliers to the remaining trajectories in terms of their topological behaviour:
First, observe that many whale trajectories are very short and only local.
They have no meaningful harmonic behaviour and are embedded very close to the origin.
Longer trajectories are located further from the origin, but still belong to the cluster of most trajectories.
They (at least partially) cycle around the Baffin Island but have no further special behaviour.
The whale $25$ is the only one that reaches far into and swims around the Hudson Bay.
Trajectories $14$ and $16$ flow in a clockwise cycle in the Hudson Strait and further up the North Atlantic Ocean --- which no other trajectory does.
In summary, the whale outlier prediction is plausible at least from a qualitative point of view.

\section{Discussion and concluding remarks}
\label{section:discussion}
We have introduced a method for outlier detection of trajectories supported on discrete spaces, based on the harmonic eigenvectors of the Hodge-Laplacian.
The resulting low dimensional representation of high-dimensional trajectories can be defined even without referring to a specific geometry or a metric space and is thus applicable also to trajectories defined, e.g., via clickstreams on the web.
Furthermore, our method naturally enables us to compare trajectories of different length without having to perform zero-padding or another form of pre-processing.

Our approach may be extended in a number of directions.
Instead of using merely the embedding of the whole trajectory (the flattened trajectory embedding), it may be useful to incorporate the incremental embeddings as well.
Furthermore, instead of relying solely on the harmonic eigenvectors of the Hodge-Laplacian, we also employ other eigenvectors of the Hodge-Laplacian associated to gradient and curl compontents~\cite{schaub2020random,schaub2021}, which would enable us to take additional features of the trajectories into account.

We have assumed above that the holes of the simplicial complex are defined a priori and can be exploited for outlier detection. 
An interesting question will be to find the optimal placement of landmarks in order to distinguish certain trajectory behaviors.
For additional flexibility in the modelling of the underlying space, we may furthermore consider cellular complexes rather than simplicial complexes~\cite{roddenberry2021signal}.

\bibliographystyle{IEEEtran}
\bibliography{bibliography.bib}

\end{document}